\documentclass[manuscript]{acmart}

\AtBeginDocument{%
  \providecommand\BibTeX{{%
    \normalfont B\kern-0.5em{\scshape i\kern-0.25em b}\kern-0.8em\TeX}}}

\copyrightyear{2022} 
\acmYear{2022} 
\setcopyright{rightsretained} 
\acmConference[MuC '22]{Mensch und Computer 2022}{September 4--7, 2022}{Darmstadt, Germany}
\acmBooktitle{Mensch und Computer 2022 (MuC '22), September 4--7, 2022, Darmstadt, Germany}\acmDOI{10.1145/3543758.3547560}
\acmISBN{978-1-4503-9690-5/22/09}

\usepackage{tikz}
\usepackage{graphicx}
\usepackage{appendix}
\usepackage{subcaption}
\usepackage{multirow}

\usepackage{hyperref}

\makeatletter
\patchcmd{\hyper@makecurrent}{%
    \ifx\Hy@param\Hy@chapterstring
        \let\Hy@param\Hy@chapapp
    \fi
}{%
    \iftoggle{inappendix}{
        \@checkappendixparam{chapter}%
        \@checkappendixparam{section}%
        \@checkappendixparam{subsection}%
        \@checkappendixparam{subsubsection}%
        \@checkappendixparam{paragraph}%
        \@checkappendixparam{subparagraph}%
    }{}%
}{}{\errmessage{failed to patch}}

\newcommand*{\@checkappendixparam}[1]{%
    \def\@checkappendixparamtmp{#1}%
    \ifx\Hy@param\@checkappendixparamtmp
        \let\Hy@param\Hy@appendixstring
    \fi
}
\makeatletter
\newtoggle{inappendix}
\togglefalse{inappendix}

\apptocmd{\appendix}{\toggletrue{inappendix}}{}{\errmessage{failed to patch}}
\apptocmd{\subappendices}{\toggletrue{inappendix}}{}{\errmessage{failed to patch}}

\makeatletter
\newcommand*{\encircled}[1]{\relax\ifmmode\mathpalette\@encircled@math{#1}\else\@encircled{#1}\fi}
\newcommand*{\@encircled@math}[2]{\@encircled{$\m@th#1#2$}}
\newcommand*{\@encircled}[1]{%
  \tikz[baseline,anchor=base]{\node[draw,circle,outer sep=0pt,inner sep=.2ex] {#1};}}
\makeatother

\begin{document}

\title[Unfolding Values through Systematic Guidance]{Unfolding Values through Systematic Guidance: Conducting a Value-Centered Participatory Workshop for a Patient-Oriented Data Donation}

\author{David Leimstädtner}
\authornote{Both authors contributed equally to this research.}
\email{david.leimstaedtner@fu-berlin.com}
\orcid{0000-0002-1445-3153}
\affiliation{%
  \institution{Freie Universität Berlin}
  \city{Berlin}
  \country{Germany}
}

\author{Peter Sörries}
\email{peter.soerries@fu-berlin.de}
\orcid{0000-0003-0493-2895}
\authornotemark[1]
\affiliation{%
  \institution{Freie Universität Berlin}
  \city{Berlin}
  \country{Germany}
}

\author{Claudia Müller-Birn}
\email{clmb@inf.fu-berlin.de}
\orcid{0000-0002-5143-1770}
\affiliation{%
  \institution{Freie Universität Berlin}
  \city{Berlin}
  \country{Germany}
}

\renewcommand{\shortauthors}{Leimstädtner and Sörries, et al.}

\begin{abstract}
Routinely collected clinical patient data posits a valuable resource for data-driven medical innovation. Such secondary data use for medical research purposes is dependent on the patient's consent.
To gain an understanding of the patients' values and needs regarding medical data donations, we developed a participatory workshop method, integrating approaches from value-sensitive and reflective design to explore patients' values and translate them into hypothetical, ideal design solutions. The data gathered in the workshop are used to derive practicable design requirements for patient-oriented data donation technologies. 
In this paper, we introduce the workshop process and evaluate its application.
\end{abstract}



\maketitle

\section{Introduction}

Nowadays, technologies supporting the disclosure, use, and sharing of clinical patient data open up promising opportunities for individualized medicine and research.
Hospitals in Germany routinely collect such data already (e.g., through health records) but their usage is often restricted to a concrete therapeutic treatment or research study. For many hospitals and researchers, it is therefore an important objective to make these data also available and reusable for secondary medical research purposes~\cite{zenker2021data}. Such secondary use, however, requires the consent of patients, which is difficult since, at the time of consent (e.g., patient admission), medical research purposes cannot be foreseen. Requesting specific consent at a later time may decrease data quality through selection bias, as patients may no longer be able to consent due to the progression of their diseases~\cite{zenker2021data}.
This situation requires a broader definition concerning the purpose of the intended data use for medical research.
In this regard, all German university hospitals agreed upon the ``broad consent'' in 2020, which defines the (re)use and storage of clinical patient data for cross-site and future medical research purposes.\footnote{For further information please visit \url{https://www.medizininformatik-initiative.de/en/template-text-patient-consent-forms}.}
Although the broad consent approach is considered feasible and legally compliant with the General Data Protection Regulation (GDPR)~\cite{zenker2021data}, it faces criticism concerning the provision of essential information (e.g., about how data are stored and shared)~\cite{grady2015broad} and restricted possibilities for withdrawing consent~\cite{mccall2018opt}.
To counteract such issues, the current version of the broad consent incorporates requirements for the secondary use of clinical patient data forwarded by stakeholder representatives such as ethics committees, data protection authorities, and a consultation group including patient representatives~\cite{hirschberg2014international, zenker2021data}.
However, selecting between the requirements of multiple stakeholders with differing interests, perspectives, and values can introduce bias through asymmetries of engagement and influence, brought about by factors like professional background, social status, or eloquence~\cite{dahl_facilitating_2020}.
These asymmetries may cause patient values to be underrepresented in terms of measures to protect and inform patients of unintentional disclosure of data and comprehensive information for transparency~\cite{zenker2021data}.

Our starting point is to examine patients' values regarding data donation and the associated context-specific, sometimes divergent attitudes, or diversity of concerns.
With our research, we assume that value-driven stakeholder participation can inform a responsible and socio-technical design of medical data donation processes.
To tackle this assumption, the research presented here is driven by the following question: \emph{How can a value-centered approach reveal values from a patient perspective and support patients' reflection on data donation processes in the medical context?}
Based on our research question, we propose a methodical approach for a value-centered participatory workshop to explore patient-oriented data donation in the medical context.
Our methodical approach is informed by two perspectives from the field of human-computer interaction (HCI): First, the theoretically founded approach \emph{value-sensitive design} (VSD)~\cite{friedman2019value}, and, second, \emph{reflective design}~\cite{sengers2005reflective}.
VSD provides methods~\cite{friedman2017methods} and toolkits (e.g., ``Envisioning Cards''~\cite{friedman2012cards}) for considering the complexity of human values within the process of technology design.
In reflective design, designers are encouraged to bring the unconscious values technology exhibits to the fore by creating technical alternatives. This approach helps designers to reflect on their own values and those of their users to incorporate them into a critical design practice~\cite{sengers2005reflective}. 

Our contribution consists of three parts: (1)~a methodical approach providing systematic guidance to conducting a value- centered participatory workshop and assisting tools to explore and reveal context-specific values, (2)~an exemplary value-centered participatory workshop carried out as a full-day experience with experts from the field of patient advocacy of vulnerable groups (e.g., patients with a rare disease), and (3) reflection on our methodical approach, taking into account the critical attitude of the participants gained through the conducted workshop, which will inform and improve future value-centered participatory workshops.
In the following, we introduce the theoretical background of our methodical approach (cf.~\autoref{sec:relatedwork}). We then detail the workshop procedure (cf.~\autoref{sec:method}), evaluate our value-centered participatory workshop (cf.~\autoref{sec:reflection}), and conclude with our planned future work (cf.~\autoref{sec:conclusion}).

\section{Leveraging Value-Sensitive Design through Reflective Design}
\label{sec:relatedwork}

VSD unites computer ethics and design methods by proposing a generative approach to define values in terms of technology design~\cite{friedman2019value}.
In VSD, human values are defined as \emph{``what is important to people in their lives, with a focus on ethics and morality''}~\cite{friedman2017methods}.
As a tripartite methodology, VSD~\cite{friedman2017methods} includes three investigations, namely \emph{conceptual}, \emph{empirical}, and \emph{technical} to address these values.
First, the \emph{conceptual} investigation defines the analytical, theoretical, or philosophical explorations of the main issues through investigation.
Second, the \emph{empirical} investigation defines the human context in which the technology is situated (e.g., through quantitative and qualitative research methods).
Third, the \emph{technical} investigation involves the analysis of the technology itself for its retrospective and future use.
VSD suggests the use of these investigations in an iterative way with the idea to improve and adjust a technology design in terms of certain stakeholder perspectives (who are directly or indirectly affected by the technology) and their values through the design process.
As a part of this tripartite methodology, VSD offers a collection of methods and toolkits that can be used in varying contexts (cf.~\cite{friedman2017methods}).
For example, Friedman and Henry's ``Envisioning cards''~\cite{friedman2012cards} raise awareness regarding the long-term and systemic effects of technologies using four criteria, namely stakeholders, time, values, and pervasiveness.
Nathan et al.~\cite{nathan2007value} introduce ``value scenarios,'' which are vignettes that are used as an analytical tool to consider values in technologies. 
In addition, Alsheri et al.~\cite{alshehri_scenario_2020} describe the use of value scenarios to envision the effect of hypothesized technology on stakeholders and their environment, emphasizing social implications. Thereby value scenarios can be used to ideate new (design) solutions.
However, VSD has been found challenging as it explores values on a universal level~\cite{friedman2002humanvalues, friedman2013value}, rather than revealing context-specific values~\cite{dantec2009experience}.

One approach to leverage VSD is \emph{reflective design}~\cite{sengers2005reflective}, which actuates designers and users to reflect on their \emph{``values, attitudes, and ways of looking at the world''}~\cite{sengers2005reflective} for the design of socially responsible technologies.
These ``ways of looking'' underpin the effort of reflective design to provoke ``critical'' reflection that aims at \emph{``bringing unconscious aspects of experience to conscious awareness''}~\cite{sengers2005reflective}.
Reflective design adopts essential approaches that are part of the third wave of HCI~\cite{harrison2011epistemological}; in addition to VSD, for example, from \emph{participatory design}~\cite{schuler1993participatory} and \emph{critical design}~\cite{dunne2001design}.
In reflective design, design itself is used to explore the needs of people in particular contexts and to guide designers and users to reflect on their practice, for example, through participatory workshops or co-creation activities.
In such activities, designers can collect and reflect on users' values and bring implicit knowledge to the fore, i.e., an ongoing inquiry in which designers are in the role of ``researchers'' situated in a ``context of practice''~\cite{schoen1983reflective}.
The boundaries and holistic implications of a technology's design are then reflected from the beginning of the design process~\cite{sengers2005reflective}.

In summary, the tripartite methodology of VSD supports flexible guidance and methods that can be incorporated into the design process. 
Nonetheless, values should be considered inductively, from the particular context in which an activity takes place, rather than assuming values and their universal properties in advance \cite{dantec2009experience}.
Reflective design embraces theoretical and methodical commitments in HCI~\cite{shilton2018values} that informed our value-centered research.

\section{Conducting the Value-Centered Participatory Workshop}
\label{sec:method}

\begin{figure*}[ht]
  \centering
  \includegraphics[width=\textwidth]{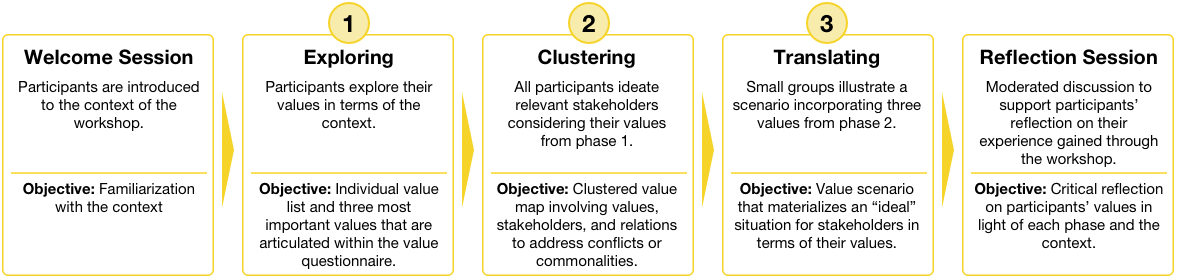}
  \caption{Procedure of the value-centered participatory workshop involving the \emph{welcome session}, phase~\protect\encircled{1}~\emph{exploring}, phase~\protect\encircled{2}~\emph{clustering}, phase~\protect\encircled{3}~\emph{translating}, and the \emph{reflection session}.}
  \Description{Procedure of the value-centered participatory workshop involving the welcome session, phase 1 exploring, phase 2 clustering, phase 3 translating, and the reflection session.}
  \label{fig:procedure}
\end{figure*}

Based on these theoretical considerations, we designed a value-centered participatory workshop consisting of a \emph{welcome session} tailored to the context of a patient-oriented data donation, the three main phases, namely \protect\encircled{1}~\emph{exploring}, \protect\encircled{2}~\emph{clustering}, \protect\encircled{3}~\emph{translating}, and a \emph{reflection session} (cf.~\autoref{fig:procedure}).
As suggested by Thabane et al.~\cite{thabane2010tutorial}, we analyzed our methodical approach (i.e., the workshop procedure) in a pilot study with six participants (four female and two male, self-reported) with an academic background (two students and four research associates).
An independent observer documented the workshop procedure that we analyzed regarding its feasibility and whether the sample size enabled appropriate collaboration and discussion among the participants.
The pilot study led to improvements in framing the workshop objective, the design of the assisting tools, and the specification of activities associated with each phase.

In May 2022, we conducted a value-centered participatory workshop (main study) in German with a group of five experts (three female and two male, self-reported). This group involved representatives from the field of patient advocacy for vulnerable groups.
The group is characterized by their personal history: participants indicated that they or a close relative suffer from a rare disease. Furthermore, a foremost critical perspective on clinical institutions was expressed based on past, emotionally draining experiences.
For the main study, we created a space where the participants felt at ease and were encouraged to share ideas and participate proactively~\cite{muller2007participatory}.
The full-day workshop (4,5 hrs, breaks excluded) was conducted in compliance with current COVID-19 hygiene protection measures.

\subsection{Workshop Procedure}

Before the welcome session, we informed participants about the purpose of the study. We carefully explained how and why workshop data (i.e., audio and video footage) is collected, stored, and processed and finally asked for their consent.
We then introduced participants to the context, purpose, and workshop objective, followed by the three previously named phases.
We concluded the workshop with a moderated reflection session.
In the following, we detail the workshop procedure.

\paragraph{Welcome Session}
We reserved time for the participants to explore the workshop materials we created, which consisted of small boxes for each participant (containing various writing and drawing utensils and sticky notes\footnote{For comparability and further analysis of this and upcoming workshops, we decided to use color-coded sticky notes, i.e., yellow for values, pink for stakeholders, and green for value conflicts or commonalities.}), accompanied by a workshop leaflet presenting information about the background, facilitators, method, and workshop program.

\begin{figure*}[t]
    \centering
    \begin{minipage}[]{\textwidth}
        \centering
        \includegraphics[width=0.325\textwidth]{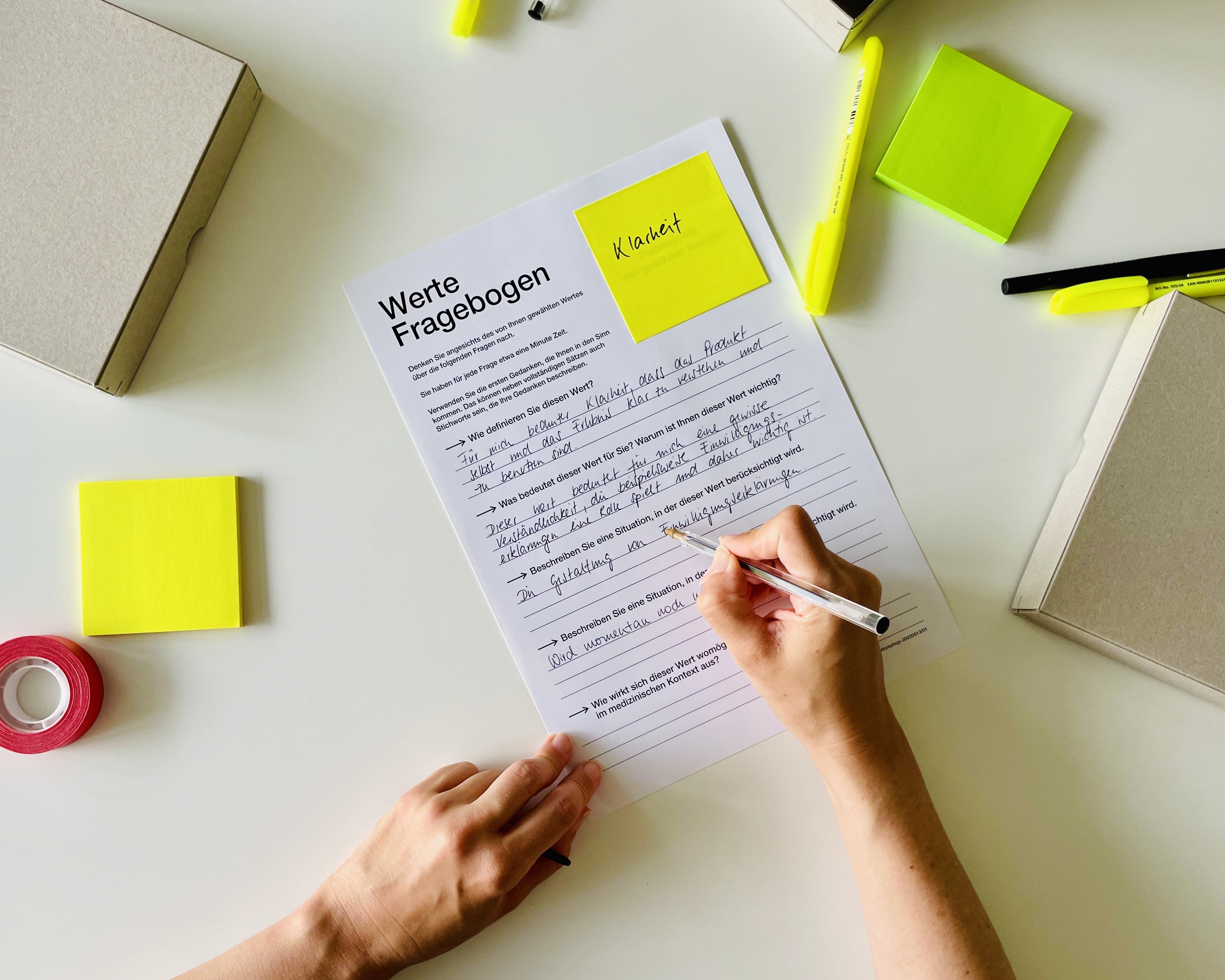}
        \hfill
        \includegraphics[width=0.325\textwidth]{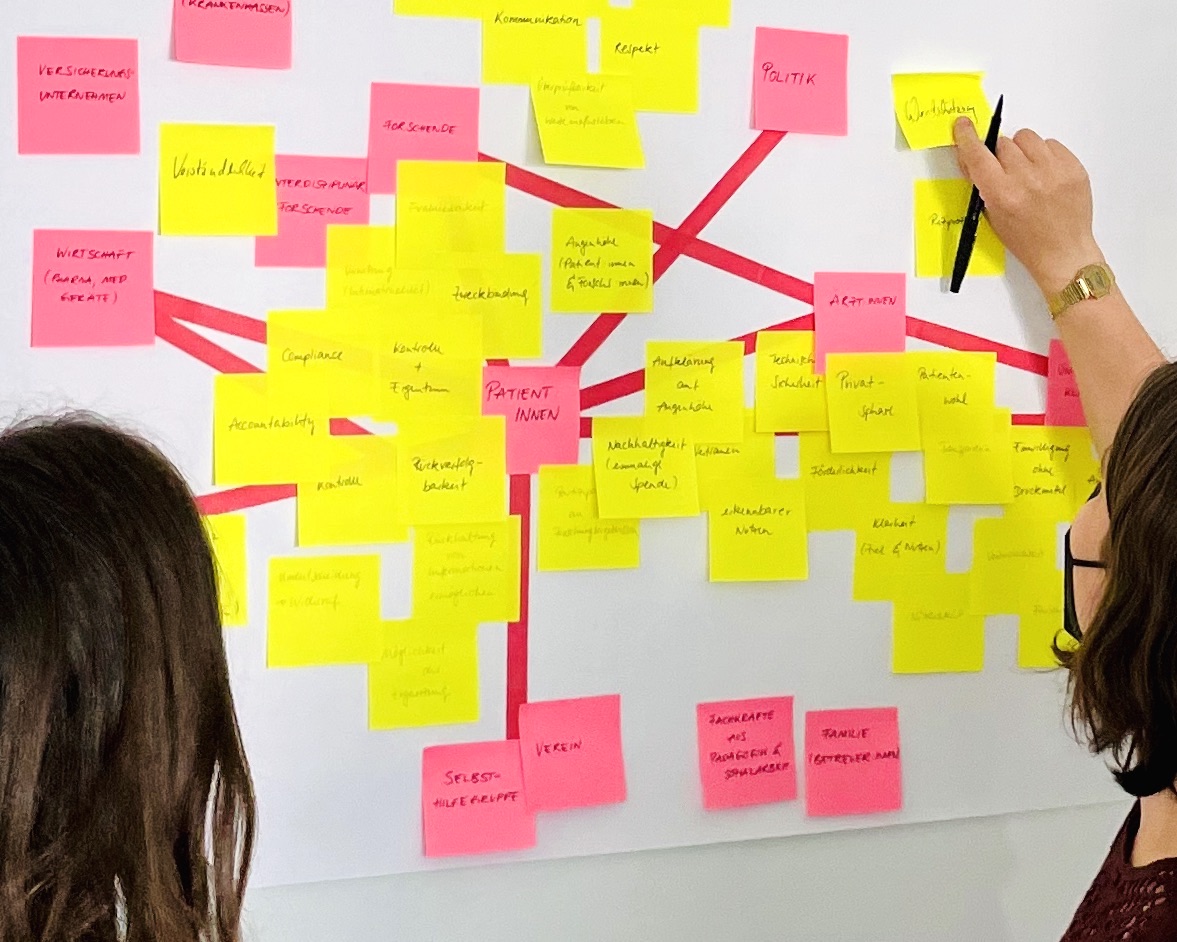}
        \hfill
        \includegraphics[width=0.325\textwidth]{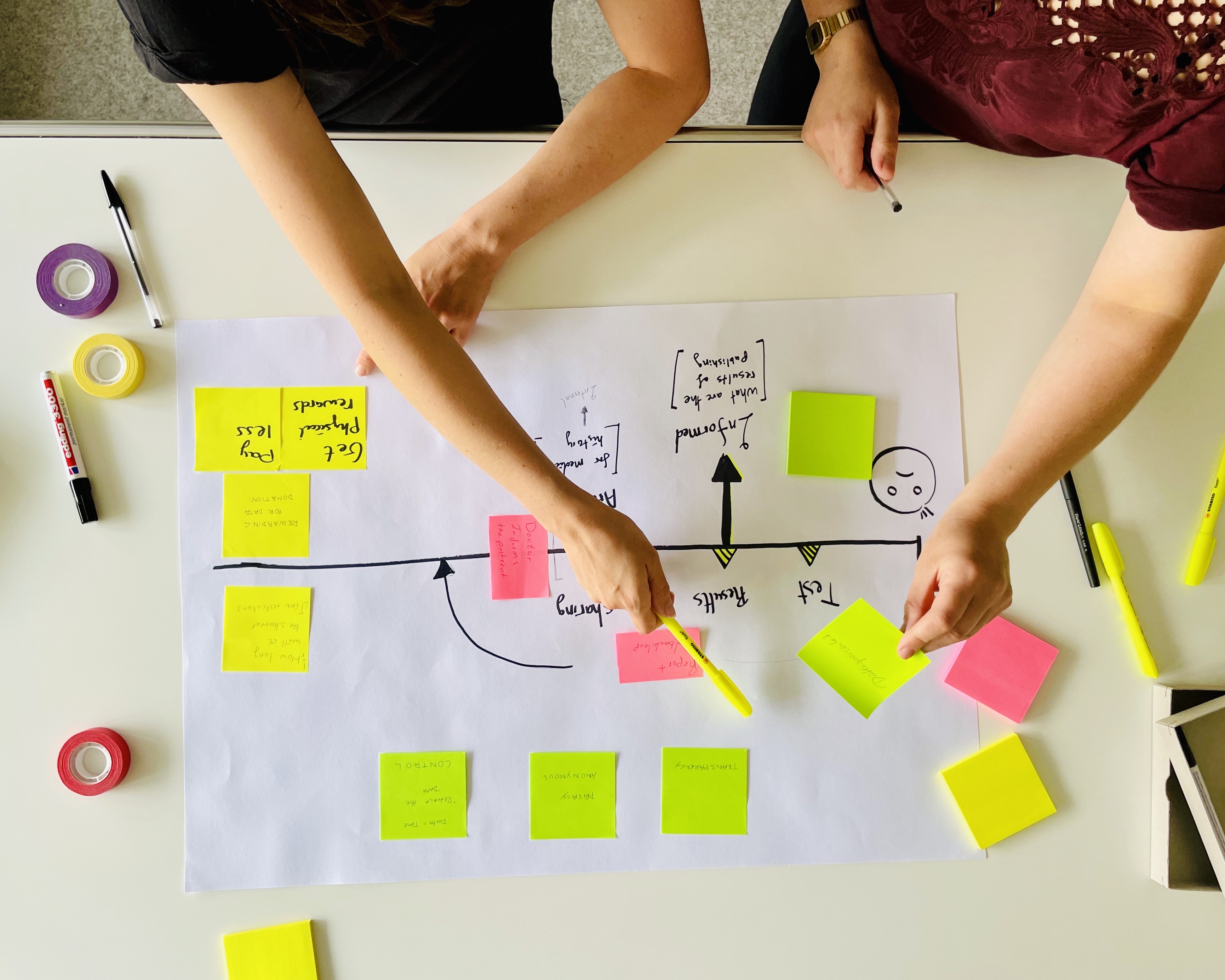}
        \caption{Assisting tools for each phase, namely the ``value questionnaire'' (phase~\protect\encircled{1}~\emph{exploring}), the ``value map'' (phase~\protect\encircled{2}~\emph{clustering}), and the ``value scenario'' (phase~\protect\encircled{3}~\emph{translating}).}
        \Description{Assisting material for each phase, namely the ``value questionnaire'' (phase 1 exploring), the ``value map'' (phase 2 clustering), and the ``value scenario'' (phase 3 translating).}
        \label{fig:material}
    \end{minipage}
\end{figure*}

\paragraph{Phase \protect\encircled{1} Exploring}
The first phase aims to elicit and understand the participants' values. 
First, to arrive at a list of situated values for the particular context~\cite{shilton2018values, dantec2009experience}, 
participants reflect on their values concerning medical data donations and write these on yellow sticky notes.
Values are frequently expressed through protean words (e.g., trust or quality)~\cite{alshehri_scenario_2020}. Therefore, we subsequently present the participants with the ``value questionnaire'' (cf.~\autoref{fig:material}, left) containing five questions (cf.~\autoref{sec:valuequestionnaire}).
Each participant selects the three most important values from the previous step to further investigate the conceptualization of their values~\cite{friedman2012cards}.
This exploratory inquiry provides participants the opportunity to articulate their understanding of a particular value (cf.~\cite{woelfer2011homeless}) and to produce a comprehensive description of their viewpoints in terms of the workshop context.
Lastly, we initiated a group discussion in which participants first selected and presented one of the three value questionnaires and then identified and commented on similarities or differences between participants' values.

\paragraph{Phase \protect\encircled{2} Clustering}
The second phase aims to cluster the collected values from phase~1 to specific stakeholders on the ``value map'' (cf.~\autoref{fig:material}, middle) to create a shared understanding of the wider context and unveil relations, commonalities, and conflicts between the stakeholders. 
The participants identify direct or indirect stakeholders (cf.~\cite{nathan2008stakeholders}) they consider to be of relevance in the context of medical data donation processes.
As an orientation, a pinboard is prepared, which includes pink sticky notes denoting the main stakeholders of interest (i.e., patients) in the center with a small selection of additional stakeholders (e.g., physicians, medical researchers) placed around the main stakeholders.
The participants are asked to add further stakeholders they consider relevant. 
Then, we guided the participants through the completion of the value map in four steps: 
First, participants place the three most important values from the previous phase on a specific stakeholder. Second, the participants rearrange stakeholders based on their assumed relation to the values. Third, participants add all further values explored in phase~1 that might be relevant, and, fourth, participants are invited to consider commonalities or conflicts, which are then marked by the facilitators using tape and green sticky notes containing explicit reasoning.
Lastly, the value map is discussed in the group to review the relevance of particular stakeholders and their values to the workshop context.

\paragraph{Phase \protect\encircled{3} Translating}
The third phase aims to render the stakeholder values into an ``ideal'' situation for data donation through the creation of value scenarios (cf.~\autoref{fig:material}, right). 
Adapting the set-up proposed by Woelfer et al.~\cite{woelfer2011homeless}, we provided the participants with a context-specific instruction (cf.~\autoref{sec:valuescenario}).
This instruction seeks to convey the interpretive flexibility of a reflective design, regarding technology design as a probe exploring values, assumptions, and needs previously excluded from design considerations~\cite{shilton2018values}.
The selected values serve thereby as hypotheticals, inspiring and guiding the exploration of new design territory within the group work~\cite{jafarinaimi2015values}.
As suggested by Alshehri et al.~\cite{alshehri_scenario_2020}, participants are invited to write, sketch, or combine both (e.g., a sketched scenario with written descriptions) when creating their value scenarios to support varying levels of confidence in sketching or writing.
The participants work in small groups (consisting of two or three participants). They are provided with large (blank) sheets and a variety of pens and sticky notes.
Lastly, each team presents its results in a ``gallery walk.'' Participants are engaged to provide feedback and discuss similarities or ambiguities between the value scenarios.

\paragraph{Reflection Session}
The reflection session aims to support participants' critical reflection.
First, the participants locate their scenario on the value map, discussing how their ideal scenario might improve the relationship between two or more stakeholders and if their values are represented appropriately.
Second, participants reflect on their stance, i.e., if a shift in the perceptions of their values has occurred throughout the three main phases.
Lastly, we conclude the workshop with a moderated feedback session in which the participants are asked to point out the contents or elements (e.g., phases or assisting tools) of the workshop they deemed most important and what considerations they missed.

\subsection{Proposed Measures for the Analysis of the Workshop}

The entirety of the workshop was recorded on audio and video and consequently transcribed verbatim.
This transcript, combined with all written material produced during the workshop, serves as data set for a thematic analysis~\cite{braun2006thematic} to identify which values and themes are most important for patient-oriented data donation.
Based on these findings, we aim to derive design requirements for technologies regarding patient-oriented data donation.
We plan to present the proposed design requirements for data donation practices based on the findings of additional workshops in a future paper since we are currently validating our methodical approach.

\section{Reflecting on the Value-Centered Participatory Workshop}
\label{sec:reflection}

According to our research question, we reflect on the feasibility of our methodical approach and if participants' critical reflection on data donation processes in the medical context was supported through the value-centered participatory workshop.
To answer our question, we focus on the findings based on data derived from a final feedback round after the workshop and an anonymous online questionnaire that participants received two weeks after the main study.\footnote{Our questions exclusively consider the three phases of the workshop and the assisting tools to validate our methodical approach (cf.~\autoref{sec:feedback}).}

In phase~\protect\encircled{1}~\emph{exploring}, we aimed to support participants' value exploration.
Participant P4 stated that VSD's understanding of values was inspiring.
However, P4 reflected: \emph{``What would have been the result if we had not seen these universal values [proposed by VSD] beforehand?''}
Referring to LeDantec et al.~\cite{dantec2009experience}, we assume that VSD's universal values presented in the welcome session unconsciously directed participants in their individual exploration of values in phase~1.
In terms of future workshops, our idea is to exemplify values in ways that are independent of a workshop context in order to represent the spectrum of values but not tamper with the actual workshop context.
Answering the value questionnaire was deemed difficult by participant P3: \emph{``[The value questionnaire] was not quite so easy. [...] I believe everyone has their own preconceived understandings of values.''}
We view such difficulty as an indication that the value questionnaire succeeded in instigating an introspective exploration of the participants' perceptions concerning their values, which have not been articulated in detail before.
In phase~\protect\encircled{2}~\emph{clustering}, we aimed to create a value map to cluster values of phase~1 to patients and additional stakeholders.
In the final feedback round, all participants emphasized the value map as a very effective vehicle to convey the stakeholders' relations and their scope of action providing an understanding of where the limits of realizing values lie.
Participant P3 underlines this: \emph{``[The value map] helped us to say at which point, which values can be adhered to [...]''}
All participants agreed on this statement and reflected that it is not necessary to include all values from phase~1 in the value map.
The value map led participants to specify a selection of central values such as awareness or traceability of data through comprehensive information.
Participant P1 suggested highlighting the patients as more prominent on the value map to constantly represent the task objective.
Participant P5 added that a more precise definition of the task is needed regarding patient values, i.e., which values should be considered for the value map (e.g., individual participant values or values related to data donation).
In phase~\protect\encircled{3}~\emph{translating}, we aimed to illustrate an ideal data donation from the patients' perspective.
Two value scenarios were created by two small groups with overlapping concepts.
Both scenarios focused specifically on the communication between patients and physicians.
Interestingly, both groups extended the task by identifying ``sub-values'' for each value to improve certain stages of a medical data donation process.
For example, group~1 (P1 and P3) ideated specific data flows and argued that their selected values (e.g., autonomy) couldn't be fulfilled equally by all stakeholders. They explained that the process of data donation is not a unidirectional process and has to consider additional values (e.g., ``control'' concerning the withdrawal of donated data at any time).
Similarly, group~2 (P2, P4, and P5) suggested a ``mediator role'' that educates about cross-site data usage.
In the reflection session, participants related the creation of value scenarios back to their inherent values.
Participant P4 stated \emph{``there is a certain ambivalence as to whether I now formulate my fundamental values or values about data donation.''}
Such ambivalence may render it difficult for participants to step outside their personal experiences and think about an ideal patient-oriented data donation in general.
This illustrates how the participants posited a very emotional view based on their personal background (e.g., medical therapies for a rare disease).

In summary, the participants perceived the procedure of the value-centered participatory workshop (the three phases and assisting tools) as logically comprehensible, which also mitigated their inner barriers to participating in each activity.
Nevertheless, we collected insights that help revise our methodical approach.
First, we will reformulate the introduction of the welcome session, i.e., move away from explicitly stating VSD's universal values to achieve freedom for participants to explore individual values.
Second, the participants inspired us to conduct further value-centered participatory workshops with physicians to assess their perspectives on patient-oriented data donation.
Participant P5 explains: \emph{``[If physicians] would now also present such a value map for themselves [...], some misunderstandings I think would also be cleared up.''}
Finally, we (facilitators) recognized the value of creating and maintaining a clearly defined ``common language''~\cite{dahl2022faciliation} for participants and us to create a deeper shared understanding such as the precise definition and reasoning of each activity.
Further, we align with research (e.g.,~\cite{nathan2007value}) that (value) scenarios unveil unforeseen aspects of a specific context that might inform novel aspects of technology design and provoke reflection; but also that further research is needed for the systematic use of VSD methods and toolkits~\cite{shilton2018values}.

\section{Conclusion and Future Work}
\label{sec:conclusion}

In our work, we leverage VSD with reflective design to create a value-centered participatory workshop divided into three phases, namely, \emph{exploring}, \emph{clustering}, and \emph{translating} with assisting tools (i.e., value questionnaire, value map, and value scenario).
As part of our ongoing research, the context of this workshop aligned with the challenge that data donation in the medical domain does not reflect patients' values.
We are currently preparing additional workshops with other stakeholders to further improve our methodical approach and collect more empirical data providing reasoning for the need to consider patient values in medical data donation practices.
We aim to derive design requirements for data donation practices based on the findings of further workshops that we present in a future paper.

\begin{acks}
We thank the participants of both workshops and the reviewers for their valuable and insightful comments. This work is supported by the Federal Ministry of Education and Research (grant 16SV8463: WerteRadar).
\end{acks}

\bibliographystyle{ACM-Reference-Format}
\bibliography{references}

\appendix

\section{Value Questionnaire}
\label{sec:valuequestionnaire}

With your chosen value in mind, please consider the following questions. You have about one minute to answer each question. Use the first thoughts that come to mind. These can be full sentences or keywords describing your thoughts.

\begin{enumerate}
\item How do you define this value?
\item What does this value mean to you? Why is this value important to you?
\item Describe a situation in which this value is given importance.
\item Describe a situation in which this value is \underline{not} taken into account.
\item How does this value possibly affect data donation in a medical context?
\end{enumerate}

\section{Value Scenario: Instruction}
\label{sec:valuescenario}

Please describe an ideal scenario for a medical data donation process from the patients' perspective. Select three values that you like to realize within the scenario. There is no right or wrong in creating the scenario. The scenario can relate to a real or fictional situation. The results will be presented to the group (five minutes each).

\section{Questions to Reflect on the Workshop}
\label{sec:feedback}

\begin{itemize}
    \item Did the exploration of values in phase~\protect\encircled{1} help you identify your fundamental values of patient-centered data donation and make values explicit through the values questionnaire?
    \item In your view, did the value map in phase~\protect\encircled{2} support you to categorize stakeholders and their values that you had not previously considered and reveal possible conflicts of values?
    \item Did the values scenario in phase~\protect\encircled{3} help you to ideate an ideal patient-oriented data donation to incorporate patients' values and expectations?
\end{itemize}

\end{document}